# Efficacy of Various Large Language Models in Generating Smart Contracts

Siddhartha Chatterjee[1] and Bina Ramamurthy[2]

[1]Mountain View High School, Mountain View, CA, USA
[2]University at Buffalo, Buffalo, NY, USA
siddhartha.chatterjee210@gmail.com

**Abstract.** This study analyzes the application of code-generating Large Language Models in the creation of immutable Solidity smart contracts on the Ethereum Blockchain. Other works have previously analyzed Artificial Intelligence code generation abilities. This paper aims to expand this to a larger scope to include programs where security and efficiency are of utmost priority such as smart contracts. The hypothesis leading into the study was that LLMs in general would have difficulty in rigorously implementing security details in the code, which was shown through our results, but surprisingly generally succeeded in many common types of contracts. We also discovered a novel way of generating smart contracts through new prompting strategies.

## 1  Introduction

Since the creation of Ethereum in 2015 which allowed for programs to run on the Blockchain, Smart Contracts have seen a surge in popularity and application. These "Smart Contracts" are really just immutable pieces of code executing directly on the Blockchain, allowing for more effective storage and exchanging of information and currency. Applications of smart contracts include business, education, and entertainment, stemming from the idea of "Web3," or a decentralized internet. With new advancements in AI in recent years such as the creation of Large Language Models (LLMs), AI has demonstrated success in generating software code across a spectrum of use cases.

However, the use of LLMs for software development has not largely extended towards the unexplored area of smart contract generation. This paper examines the efficacy of various LLMs in this specialized, yet crucial, task. The importance of such a study is highlighted by the critical need for auditing smart contracts as, once published, they are unalterable, necessitating optimal security and efficiency before being appended to the Blockchain. Any vulnerabilities can open the door to hacking incidents, leading to substantial losses of money and data, such as has been the case with numerous Web3 companies in the past. Given the rising usage of AI in code generation, a pressing question emerges: can AI-generated code meet the stringent security standards required in smart contracts? This research hopes to provide insights into this crucial query.

## 1.1 Background

The study will aim to answer the following research questions:

- How accurately can different LLMs generate smart contracts?
- What are the specific strengths and weaknesses of each evaluated LLM in the context of smart contract generation?
- How do LLM-generated smart contracts compare with manually created contracts in terms of reliability, accuracy (Mastropaolo et al. [10]), safety and efficiency?

## 1.2 Document Organization

The document is organized as follows:

- **Section 2 (Related Work)**: Reviews research on AI code generation, focusing on program synthesis and smart contracts, particularly advances in using Large Language Models (LLMs).
- **Section 3 (Methodology)**: Describes the experimental setup for evaluating LLMs in generating Solidity smart contracts, including model selection and assessment criteria.
- **Section 4 (Results)**: Presents findings on LLM performance in smart contract generation, comparing functionality, efficiency, and code quality across models.
- **Section 5 (Conclusion)**: Summarizes key insights and proposes future research directions in blockchain exploration and LLM prompting strategies.

## 2  Related Work

Generative Artificial Intelligence (also referred to as Gen AI) [1] [2] [3] [4] is opening the doors for new ways to write programs [23] where humans and machines collaborate to various degrees. We've seen progress in two popular directions: program induction and program synthesis. Program induction is the process of automatically generating computer programs from a set of input-output examples or specifications. "Learning to Execute" by Wojciech Zaremba and Ilya Sutskever (2014) [29] explores the capabilities of Long Short-Term Memory (LSTM) networks in learning to execute simple programs by treating the problem as a sequence-to-sequence task. The paper "Neural Turing Machines" by Graves et al. (2014) [21] introduces a novel neural network architecture that combines the learning capabilities of neural networks with the storage and algorithmic power of Turing machines. The model features a neural network controller paired with an external memory bank, enabling it to perform complex tasks like algorithmic operations, sequence processing, and learning simple programs by reading from and writing to memory. The paper "Neural Programmer-Interpreters" by Reed & de Freitas (2016) [30] explores the use of neural networks to execute programs by interpreting code as data, leveraging techniques from machine learning to improve program synthesis and execution. This approach enhances the generalization capabilities of neural networks in understanding and generating code, with applications in automating complex programming tasks and improving AI-driven software development. The paper "Evaluating Large Language Models Trained on Code" [5] assesses the performance of language models specifically trained on programming languages, focusing on their ability to understand, generate, and complete code. It highlights the models' strengths in handling complex coding tasks, suggesting their potential to assist in software development and improve coding efficiency. The paper "Neural Programmer-Interpreters" additionally highlighted the ability of machines to perform reasoning tasks to create an input-output mechanism [18]; such can be applied to generating code from a prompt. In the paper, "An Empirical Evaluation of GitHub Copilot's Code Suggestions" authors presented an evaluation of GitHub Copilot. The authors conducted a study to assess the accuracy, efficiency, and quality of Copilot's suggestion [9]. The paper "Evaluating Large Language Models Trained on Code" also concluded that Copilot excelled at simple generation tasks [20]. Another study, "Asleep at the Keyboard? Assessing the Security of GitHub Copilot's Code Contributions" found Copilot generated code often featured security vulnerabilities [7]. Our study mirrors this process but with the generation of Smart Contracts instead.

  Program synthesis is the process of automatically creating a computer program that meets a given high-level specification or set of input-output examples [14]. In program synthesis, models explicitly generate programs from natural language specifications, a task crucial for automating code generation. The paper "NAPS: Natural Program Synthesis Dataset" has set a benchmark for natural program synthesis by introducing a large common dataset of natural language patterns to use to train models [38]. One classical approach in program synthesis employs a probabilistic context-free grammar (PCFG) to construct a program's abstract syntax tree (AST) (Bodík et al., 2013) [33]. Maddison and Tarlow (2014) [31] improved this approach by introducing a state vector to condition child node expansion, enhancing the generative process. This idea was later adapted by Allamanis et al. (2015) [32] for text-to-code retrieval, demonstrating its effectiveness in matching code snippets to textual queries. Yin and Neubig (2017) [34] further extended this concept to text-conditional code generation, enabling the direct generation of code from natural language descriptions. In the paper "On Domain Knowledge and Novelty to Improve Program Synthesis Performance with Grammatical Evolution", Hemberg, Kelly, and O'Reilly [15] explored methods to enhance the efficiency of program synthesis using Grammatical Evolution (GE). The study incorporates domain knowledge to guide the design of grammar rules and introduces novelty search strategies to balance exploration and exploitation during the evolutionary process. More recent examples include using genetic algorithms, which draws on evolutionary ideas to develop complex tree-based programs, synthesized in the paper "A Representation for the Adaptive Generation of Simple Sequential Programs" [16]. Programs can be synthesized without relying on an abstract syntax tree (AST) representation, leveraging various alternative approaches. This was also shown in the paper "Recent Developments in Program Synthesis with Evolutionary Algorithms" by Dominik

Sobania, Dirk Schweim, and Franz Rothlauf [13]. Hindle et al. (2012) [42] explored n-gram language models for code, discovering that code is significantly more predictable than natural language due to its structured nature. This predictability was further exploited by Hellendoorn and Devanbu (2017) [45], who enhanced n-gram models with caching mechanisms to improve code completion tasks. Ling et al. (2016) [46] introduced Latent Predictor Networks, demonstrating that character-level language models could generate functional code for implementing Magic the Gathering cards in an online arena by using a latent mode to copy card attributes directly into the code. This approach highlights the potential of character-level models for complex code generation tasks. DeepCoder (Balog et al., 2017) [39] advanced this field by training a model to predict the functions that appear in source code, effectively guiding the program synthesis process through a search mechanism. Rabinovich et al. (2017) [36] and Brockschmidt et al. (2020) [40] introduced graph-based neural networks to model the structural dependencies in code, bypassing the need for explicit AST representations and enabling more flexible program synthesis. These innovations illustrate the diverse methodologies being explored to synthesize programs directly from raw data, underscoring the evolving landscape of automated code generation (Sun et al., 2020 [41]). The paper "Neuro-Symbolic Program Synthesis" discussed a hybrid approach to program synthesis by combining neural networks and symbolic reasoning to perform complex reasoning tasks [35]. Another attempt at improving program synthesis was analyzed in the paper "Program Synthesis with Learned Code Idioms" whereby authors evaluated using recognizable code patterns in generating programs [37].

The area of smart contract generation using LLMs is relatively new - as is the whole field of smart contracts itself. In the paper "Who is Smarter? An Empirical Study of AI-Based Smart Contract Creation", Karanjai et. al [26], evaluated the effectiveness of AI systems in generating smart contracts as compared with human-written equivalents. The study compared various AI models on their ability to create secure and functional smart contracts, highlighting strengths and weaknesses in different approaches. Results indicated that while AI can significantly aid in smart contract development, there were still limitations in terms of security and reliability that need to be addressed. In the paper "Automatic Smart Contract Comment Generation via Large Language Models and In-Context Learning" by Zhao et. al. [11] proposes SCCLLM, a framework that employs large language models with an in-context learning approach to generate comments for smart contract code. The paper "The Hitchhiker's Guide to Program Analysis: A Journey with Large Language Models" by Li et al. (2023) [19] investigates the role of LLMs in analyzing program behavior, demonstrating how they can be used to understand and identify code vulnerabilities and improve program comprehension. The study explored how domain-specific fine-tuning and prompt engineering can enhance LLMs' ability to accurately analyze and generate explanations for different segments of code - a useful resource for debugging code. The paper "Large Language Model-Powered Smart Contract Vulnerability Detection: New Perspectives" by Sihao et. al. (2023) [43] explores the application of large language models (LLMs) to identify vulnerabilities in smart contracts. By leveraging the advanced natural language processing capabilities of LLMs, the study demonstrates significant improvements in detecting security flaws and enhancing the reliability of smart contracts. In the paper

"Combining Fine-Tuning and LLM-based Agents for Intuitive Smart Contract Auditing with Justifications", Ma et. al. [2024] [44] describes limitations of LLMs in auditing smart contracts and proposes an agentic approach with two LLMs that demonstrated significant improvement over using a single LLM. The paper "Leveraging Fine-Tuned Language Models for Efficient and Accurate Smart Contract Auditing" by Wei et. al. (2024) [28] introduces a framework that enhances vulnerability detection in smart contracts by fine-tuning LLMs with advanced prompt engineering, demonstrating superior accuracy and efficiency over traditional auditing tools. In the paper "Automatic smart contract comment generation via large language models and in-context learning", Zhao et. al. (2024) [11] describes using in-context learning to generate comments in smart contract programs.

## 3 Methodology

To study the application of AI in Smart Contract generation, we will select a range of representative Large Language Models (LLMs) that are widely used or have shown potential in the generation of smart contracts: GPT 3.5, GPT 4, GPT 4-o, Cohere, Mistral, Gemini and Claude. We'll use these models to generate smart contracts and then evaluate those smart contracts for accuracy, efficiency and code quality to benchmark our results.

We will use the following sequence of steps for each contract:

1. Prompting Techniques: We will have both descriptive prompts and structured prompts to feed LLMs. They will outline exactly the variables and functions needed. It is necessary for all prompts to contain some ambiguity in relation to the methods or style of writing for writing the Smart Contract - this is needed to test the LLM abilities.

    a. The descriptive prompt mimics how an average user may prompt the LLM, describing it while the structured prompt provides a complete outline for the contract.

    b. The structured prompt will be similar to pseudo-code. It is expected the LLMs will perform better with structured prompts, but we will primarily analyze results on descriptive prompts as that is what is most commonly used when generating code from LLMs.

2. Generation of Smart Contracts: We will provide each LLM with the same set of prompts (descriptive prompts that outline exactly the variables and functions needed), designed to create a variety of smart contracts.

3. Testing: We will write a test file in TypeScript to evaluate the performance of the contracts through a series of test cases that the code must pass. The test file will be written using the HardHat environment library allowing us to run all Smart Contract programs in a controlled Ethereum environment.

4. We'll evaluate each contract on multiple dimensions such as:

    ○ Accuracy - How many tests the code passes

    ○ Efficiency - How efficient was the code in terms of execution time

    ○ Verboseness/Quality - Quality of code outputted (qualitative)

        ■ Including potential security risks

Our results follow as a series of different applications of smart contracts of ranging complexities and the performance of each AI (compiled in a table at the end together). We've chosen a few commonly recurring smart contract patterns to create benchmark results.

### 3.1    Reading and writing a variable onto the Blockchain

**Overview:**
This is the most basic type of smart contract which is to simply store and edit a variable on the Ethereum Blockchain. It is expected that all LLMs should be able to complete this task fully.

**Descriptive Prompting:**

| Create a smart contract using solidity 0.8.20 called "Variable" with a strictly positive, public variable "val" that is defined in the constructor with an input parameter, a function "modify" with a parameter that the contract sets "val" to and a function "retrieve" that returns the value of the variable |
|---|

**Structured Prompting:**

| contract Variable |
|---|
| +      val: uint public |
| +      constructor(uint _val) |
|          // Sets var to parameter _val |
| +      modify(uint _val) |
|          // Sets var to parameter _val |
| +       retrieve() returns uint |

```
    // return value of val
```

**Test Cases:**
1. Set variable correctly after initializing contract
2. Retrieve function returns variable correctly
3. Modify fails if negative is passed into function
4. Modify changes variable
5. Retrieve function returns variable correctly after modification

**General Observations for Case 1**
GPT-3.5 added an unnecessary check of the variable to be greater than 0, which is not necessary since the data type is "uint" and will revert automatically if the parameter is negative

```
 // GPT 3.5
constructor(uint256 initialValue) {
    require(initialValue > 0, "Initial value must be strictly positive");
    val = initialValue;
}
```

Cohere didn't include an SPDX license identifier which resulted in a warning

```
// Cohere Coral
pragma solidity 0.8.20;

contract Variable {
```

### 3.2 Lock some amount of money for finite amount of time on the Ethereum Blockchain

**Overview:**
This contract is developed by Hardhat (a Web3 programming environment) in their initialization example. It is relatively simple and not meant to be used exactly on scale, but it is useful to see how the LLMs will perform on this initial straightforward task.

**Descriptive Prompt:**

8  Chatterjee, et al.

> Create a smart contract using Solidity 0.8.20 called "Lock" with public variables "unlockTime" that gets defined in the constructor by the parameter and "owner" which is set to the address that initializes the contract, it also receives some amount of ether which is stored in the contract. Write a function "withdraw" that sends the balance of the contract to the address upon the conditions that the time is passed the unlockTime and the address is the owner address. This function should trigger an event "Withdrawal" with parameters of the balance and the timestamp

**Structured Prompting Technique:**

| contract Lock |
| --- |
|     +    unlockTime: int |
|     +    owner: address, payable |
|     +    event Withdrawal(uint amount, uint time) |
|     +    constructor(uint unlockTime) |
|         // initializes owner and unlockTime |
|     +    withdraw() |
|         // sends money if criteria is met |

**Test Cases:**
1. Set correct unlockTime
2. Set correct owner
3. Receive funds and store in contract
4. Revert if set unlockTime is below current time (in the past)
5. Revert if withdraw is called before unlockTime
6. Revert if account calling withdraw is not owner
7. Should not fail if account is owner and unlockTime has passed
8. Emit event after successful withdrawal
9. Send funds to owner account after successful withdrawal

**General Observations for Case 2**

GPT 3.5 added extra parameter of the receiver (not specified in either the descriptive or structured prompt) for Withdrawal (unspecified and unnecessary):

```
// GPT 3.5

event Withdrawal(address indexed receiver, uint256 amount, uint256 timestamp);
```

Additionally GPT 3.5 didn't make the constructor payable (should have been assumed from the prompt, which did not allow it to function at all).

```
// GPT 3.5

constructor(uint _unlockTime) {
    unlockTime = _unlockTime;
    owner = payable(msg.sender);
}
```

Bard included verbose "public" modifier for constructor, which is unnecessary

```
// Bard

constructor(uint256 _unlockTime) public payable {
    unlockTime = _unlockTime;
    owner = msg.sender;
}
```

Both GPT-4 and Bard overlooked the obvious test case that the unlockTime must be greater than the initialization time, and failed that test case (potentially could have lead to a bug or security issue)

```
/ GPT 4
  constructor(uint _unlockTime) payable {
      owner = msg.sender;
```

```
    unlockTime = _unlockTime;
  }
```

Mistral did not produce compilable code, with a syntax error trying to access balance before contract definition.

Cohere added an extra parameter to "Withdrawal"

```
// Cohere

event Withdrawal(address recipient, uint amount, uint timestamp);
```

### 3.3  Create a new token on Ethereum Blockchain

**Overview**

Next, we examine LLMs in a much more practical use case, the creation of a standardized ERC20 coin with custom features. We can customize the coin with

1. Supply
2. User authentication and class of User
3. Storage, exchanging, and payment methods (such as allowances from the contract)

We embedded these customizations into our prompt to add complexity.

**Descriptive Prompt:**

Create a ERC 20 Crypto token using Solidity 0.8.20 with following properties:
- Name of the token is "LLM Token"
- We'll allow 2 decimal places
- Require users to register/sign up
- Minting not allowed till certain date/time inputted into contract constructor

- Allow users to mint new tokens by depositing some ether
- The token has lazy supply starting at 10M
- Rank for different users (as a data structure)
  - 0: can only trade/swap
  - 1: can mint/burn
- Users who are rank 1 can grant other users rank 1

**Structured Prompt:**

| contract LLMToken (ERC 721) |
| --- |
| +     struct User: |
|         // Defined object with name, address, rank |
| +     usersList: User[] |
| +     mintingTime: uint |
| +     constructor() |
|         // set minting time, initialize ERC721 token with supply at 10M |
| +     mint() |
|         // mint new token, only allowed for rank 1 |
| +     burn(token) |
|         // burn token, only allowed for rank 1 |
| +     transfer(address) |
|         // transfer tokens between users, allowed for all rank |
| +     grantRank(address) |
|         // grants rank 1 to address only if rank of user is |

**Test Cases**
1. Create supply of tokens
2. Allow user to register
3. Allow only rank 1 user to mint
4. Transfer from two accounts
5. Allow rank 1 user to grant rank 1 to another user

**General Observations for Case 3**

Using imported libraries for contract, GPT 4 imported openZeppelin to use the standardized ERC 20 token. This is significant as openZeppelin is a library that is commonly the standard for ERC 20 tokens in industry nowadays.

```
// GPT 4
// SPDX-License-Identifier: MIT

pragma solidity ^0.8.20;

import "@openzeppelin/contracts/token/ERC20/ERC20.sol";
import "@openzeppelin/contracts/access/Ownable.sol";
```

This is the first case where the AI models noticeably start to differentiate as neither GPT 3.5 nor Gemini Pro implemented the openZeppelin libraries by default. This is inefficient to a large degree, as the contracts necessary for ERC are already built in but these models try to do it themselves. Additionally GPT 3.5 uses the wrong version of Solidity specified.

```
// GPT 3.5
// SPDX-License-Identifier: MIT
pragma solidity ^0.8.0;

contract LLMToken {
```

```
// Gemini
pragma solidity ^0.8.20;

// Standard ERC20 token with basic functionalities
```

```
abstract contract ERC20 {
    function transfer(address recipient, uint256 amount) public virtual;
    function balanceOf(address account) public view virtual returns (uint256);
    function allowance(address owner, address spender) public view virtual returns (uint256);
    function approve(address spender, uint256 amount) public virtual;
    function transferFrom(address sender, address recipient, uint256 amount) public virtual;
    event Transfer(address indexed from, address indexed to, uint256 value);
    event Approval(address indexed owner, address indexed spender, uint256 value);
}
```

However, while GPT-4 recognized the use of common libraries such as Openzeppelin, the code was outdated and did not initialize the implemented contracts which caused a compilation error leading to the code not being able to be run.

```
constructor(uint256 _mintStartTime) ERC20("LLM Token", "LLM") {
    require(_mintStartTime > block.timestamp, "Mint start time must be in the future");
    mintStartTime = _mintStartTime;
    _mint(msg.sender, INITIAL_SUPPLY); // Mint initial supply to contract owner
}
```

## 4 Results

### 4.1 Results Using Text-Based Prompt

Table 1: Comparing accuracy smart contract code generated by public LLMs using Text-based prompting strategy

| Tests | GPT 4 | GPT 4-o | GPT 3.5 | Cohere | Mistral | Claude | Gemini |
|---|---|---|---|---|---|---|---|
| Storage | 100% | 100% | 100% | 100% | 100% | 100% | 100% |
| Lock | 89% | 100% | 0% | 78% | N/A | 100% | 89% |
| Create Token | N/A | N/A | N/A | N/A | N/A | N/A | N/A |

$$(1)$$

**Table 2: Comparing execution efficiency Smart Contract code generated by public LLMs using Text-based prompting strategy**

| Tests | GPT 4 | GPT 4-o | GPT 3.5 | Cohere | Mistral | Claude | Gemini |
|---|---|---|---|---|---|---|---|
| Storage | 2s | 2s | 2s | 2s | 2s | 2s | 2s |
| Lock | 2s | 2s | N/A | 3s | N/A | 3s | 2s |
| Create a New Token | N/A | N/A | N/A | N/A | N/A | N/A | N/A |

(2)

**Table 3: Comparing quality of Smart Contract code generated by public LLMs using Text-based prompting strategy**

| Tests | GPT 4 | GPT 4-o | GPT 3.5 | Cohere | Mistral | Claude | Gemini |
|---|---|---|---|---|---|---|---|
| Storage | Excellent | Excellent | Unnecessary variable check | No SPDX license identifier (warning flag) | Excellent | Excellent | Excellent |
| Lock | Overlook of input check | Excellent | Extra unspecified parameter, did not make constructor payable | Extra unspecified parameter | Did not compile | Excellent | Overlook of input check |
| Create a New Token | Missing some features, did not compile | Majority code is excellent but few syntax errors | Missing some features, did not compile | Missing some features, did not compile | Did not compile | Did not compile | Majority code is excellent but few syntax errors |

(3)

## 4.2  Results Using Structured Prompt

**Table 4: Comparing accuracy smart contract code generated by public LLMs using structured prompting strategy**

| Tests | GPT 4 | GPT 4-o | GPT 3.5 | Cohere | Mistral | Claude | Gemini |
|---|---|---|---|---|---|---|---|
| Storage | 100% | 100% | 100% | 100% | 100% | 100% | 100% |
| Lock | 89% | 100% | 0% | 89% | N/A | 0% | N/A |
| Create a New Token | N/A | N/A | N/A | N/A | N/A | N/A | N/A |

(4)

**Table 5: Comparing execution efficiency Smart Contract code generated by public LLMs using structured prompting strategy**

| Tests | GPT 4 | GPT 4-o | GPT 3.5 | Cohere | Mistral | Claude | Gemini |
|---|---|---|---|---|---|---|---|
| Storage | 2s | 2s | 2s | 2s | 2s | 2s | 2s |
| Lock | 3s | 2s | N/A | 3s | N/A | N/A | N/A |
| Create a New Token | N/A | N/A | N/A | N/A | N/A | N/A | N/A |

(5)

**Table 6: Comparing quality of Smart Contract code generated by public LLMs using structured prompting strategy**

| Tests | GPT 4 | GPT 4-o | GPT 3.5 | Cohere | Mistral | Claude | Gemini Pro |
|---|---|---|---|---|---|---|---|
| Storage | Excellent | Excellent | Excellent | Excellent | Excellent | Excellent | Excellent |
| Lock | Excellent | Excellent | Did not make contract payable, could not pass any test case | Excellent | Compilation error, used nonsensical modifier "public payable" | Did not make contract payable, could not pass any test case | Compilation error, tried to convert between two non-convertible data types |
| Create a New Token | Missing some features, did not compile | Majority code is excellent but few syntax errors | Missing some features, did not compile | Missing some features, did not compile | Missing many features, Did not compile | Did not compile | Majority code is excellent but few syntax errors |

(6)

### 4.3   Significance of Differing Prompting Strategies

From our experiments, we find that foundational models generally performed better using descriptive prompting as opposed to structured prompting. This is likely because these models have been trained more extensively with human-like input, which is closer to descriptive prompts. Descriptive prompts provide natural language instructions that align more closely with the models' training data, resulting in better understanding and generation of code. However, structured prompting, which involves providing more detailed and specific instructions akin to pseudo-code, often resulted in lower performance not due to code inefficiency but primarily because of compilation errors. This suggests that while structured prompts hold potential for precise code generation, the models require further training to handle the specificity and technical demands of such prompts effectively.

Models like Cohere demonstrated an ability to adapt to structured prompting and achieved more success using this method, indicating that with sufficient training and adaptation, structured prompting could become more viable. In the future, we believe that structured prompting will become the norm for code generation as models are further trained on extensive libraries that include both natural language descriptions and structured instructions. For now, using human-like descriptions as opposed to pseudo-code remains the optimal strategy for generating accurate code, particularly for applications such as smart contract development.

### 4.4   **Overall Findings**

From the qualitative and quantitative data collected, we can determine several key insights about the performance of large language models (LLMs) in smart contract generation:

1. **Best Performers:** Claude and GPT-4-o emerged as the best performers in our evaluations. These models demonstrated superior capabilities in generating accurate and functional smart contracts compared to other models tested.

2. **GPT-4 vs. GPT-3.5:** GPT-4 significantly outperformed GPT-3.5 in terms of code accuracy and overall performance. In particular, notice Table 1 & Table 4 where GPT-4 performed at an 89% accuracy for "Lock" compared to 0% for GPT-3.5 The advancements in GPT-4 over its predecessor are evident in its ability to handle complex tasks more effectively and generate more reliable code.

3. **GPT-4-o Improvements:** GPT-4-o represents an improvement over previous models, particularly in generating accurate smart contracts. The enhancements in this model contribute to better handling of the specific requirements of smart contract coding.

4. **Code Quality Issues:** Despite the improvements, the quality of code generated by LLMs is still lacking. Many instances of redundant code, outdated code, and compilation errors were observed, especially with GPT-3.5. This highlights the need for further refinement and training to improve code generation quality (see Table 3 & Table 6). Code was also rather inefficient compared to human-written equivalent contracts.

5. **Security Overlooked:** Except for Claude, all models tended to overlook security issues unless explicitly mentioned in the prompt. This indicates a critical area where LLMs need improvement to ensure the generation of secure smart contracts. For example, in Case 2, both Bard and GPT 4-o had a vulnerability in which it allowed an "unlock time" to be set earlier than the current time of generation. This could be exploited by programs that can alter times to invalidate transfers. This is a massive security risk and underscores one of the main reasons LLMs are not viable for industry use in smart contract generation currently.

6. **Inconsistency in Complex Tasks:** For more complex tasks, all models exhibited inconsistency, indicating that while they can handle simpler tasks, they struggle with the complexity and intricacies of more advanced coding requirements. Notice Table 2 & Table 5, none of the LLMs were able to succeed in the complex task of generating an ERC-721 token.

## 5   Conclusion

Overall, our study evaluated the capabilities of Large Language Models (LLMs) in generating Solidity-based smart contracts running on the Ethereum network, focusing on security, efficiency, and accuracy. Among the models tested, advanced versions like GPT-4-o and Claude performed the best, demonstrating strong results with minimal errors in simpler contract generation. However, all models, including these leaders, exhibited inconsistencies when dealing with more complex contracts (Table 2 & Table 5).

One of the most critical limitations identified was in security implementation—a vital aspect for blockchain applications [22] [27]. Earlier models, such as GPT-3.5, often generated redundant or outdated code and encountered frequent compilation errors. Interestingly, structured prompts, despite being designed for precision, posed more challenges than descriptive ones across all models. The efficiency of the programs is noticeably slow currently, which is a huge problem with large Ethereum gas fees posing large issues for Smart Contract developers [6].

However, despite these concerns, LLMs showed promise when it came to less complicated smart contract generation. This suggests the potential for LLMs to be used to adapt existing written smart contracts for a wider application of purposes. For example, a developer may want to create a program that uses an algorithm existing in other smart contract-based apps and just needs to be re-applied to the program's specific circumstances. The potential in this area also suggests LLMs may, in the future, be able to become more precise and secure smart contracts in the future.

Currently, we can confidently say LLMs are not at a viable point to be used for smart contract generation for industrial use.

## 6  Future Work

As we've discussed, moving to more structured prompts significantly improves the quality and efficiency of smart contracts. We'll continue to explore other innovative prompting techniques that can further improve smart contract quality.

In the current research we've focused only on Ethereum Blockchain which is the most popular Blockchain amongst developers. This means there are more examples available for LLMs to learn from. We intend to extend this work for other Blockchains like Solana, Aptos, Sui etc and test the ability of LLMs to generalize code generation across different Blockchains.

Another important aspect of future research will be exploring how valuable human smart contract developers find the generated code. In the paper 'Expectation vs. Experience: Evaluating the Usability of Code Generation Tools Powered by Large Language Models', Vaithilingam, Zhang, and Glassman [8] explored the effectiveness and user perceptions of AI-powered code generation tools like GitHub Copilot. We plan to conduct similar studies with groups of smart contract developers to understand their perceptions of the usability of the generated code. Evaluating the ability of large language models to understand code syntax will also be another area of focus for us in the future. In the paper "Benchmarking Language Models for Code Syntax Understanding", Shen et. al. [17] described a method to evaluate how well leading pre-trained models such as CuBERT and CodeBERT comprehend code structure. We plan to extend their methodology to the area of smart contract generation.

Furthermore, the security of smart contracts remains a critical concern, as numerous instances of smart contracts being exploited by malicious actors have highlighted significant vulnerabilities. To address this, we plan to conduct a comprehensive investigation into the security aspects of the generated smart contracts. This will involve integrating advanced security analysis tools and techniques into our evaluation framework, aiming to identify and mitigate potential vulnerabilities proactively. By doing so, we hope to enhance the overall reliability and safety of AI-generated smart contracts, making them more resilient against exploitation. Studies such as "Combining Fine-Tuning and LLM-based Agents for Intuitive Smart Contract Auditing with Justifications" has shown the potential in LLMs auditing Smart Contracts [12]. This skill may prove more valuable to developers than full contract generation. Performance of code may also be enhanced by code-analyzing LLMs [24] [25].

Additionally, we will explore the integration of formal verification methods to rigorously prove the correctness and security of smart contracts generated by LLMs. This multidisciplinary approach, combining AI, blockchain technology, and formal methods, has the potential to set new standards for smart contract development and deployment, ensuring higher levels of trust and reliability in decentralized applications.

20      Chatterjee, et al.